\documentclass[aps,floats]{revtex4}
\usepackage{amsmath,amssymb}
\usepackage{graphicx,epsfig}

\begin{document}
\bibliographystyle {plain}

\def\oppropto{\mathop{\propto}} 
\def\opmin{\mathop{\min}} 
\def\opmax{\mathop{\max}} 
\def\opsimeq{\mathop{\simeq}}
\def\opoverderline{\mathop{\overline}}
\def\operarrow{\mathop{\longrightarrow}}
\def\opsim{\mathop{\sim}}

\def\fig#1#2{\includegraphics[height=#1]{#2}}
\def\figx#1#2{\includegraphics[width=#1]{#2}}


\title{ Statistics of renormalized on-site energies and renormalized hoppings \\
for Anderson localization models in dimensions $d=2$ and $d=3$ } 


 \author{ C\'ecile Monthus and Thomas Garel }
\affiliation{Institut de Physique Th\'{e}orique, CNRS and CEA Saclay
 91191 Gif-sur-Yvette cedex, France}

\begin{abstract}
For Anderson localization models, there exists an exact real-space renormalization procedure at fixed energy which preserves the Green functions of the remaining sites [H. Aoki, J. Phys. C13, 3369 (1980)]. Using this procedure for the Anderson tight-binding model in dimensions $d=2,3$, we study numerically the statistical properties of the renormalized on-site energies $\epsilon$ and of the renormalized hoppings $V$ as a function of the linear size $L$. We find that the renormalized on-site energies $\epsilon$ remain finite in the localized phase in $d=2,3$ and at criticality ($d=3$), with a finite density at $\epsilon=0$ and a power-law decay $1/\epsilon^2$ at large $\vert \epsilon \vert$. For the renormalized hoppings in the localized phase, we find: ${\rm ln} \ V_L \simeq -\frac{L}{\xi_{loc}}+L^{\omega}u$, where $\xi_{loc}$ is the localization length and $u$ a random variable of order one. The exponent $\omega$ is the droplet exponent characterizing the strong disorder phase of the directed polymer in a random medium of dimension $1+(d-1)$, with $\omega(d=2)=1/3$ and $\omega(d=3) \simeq 0.24$. At criticality $(d=3)$, the statistics of renormalized hoppings $V$ is multifractal, in direct correspondence with the multifractality of individual eigenstates and of two-point transmissions. In particular, we measure $\rho_{typ}\simeq 1$ for the exponent governing the typical decay $\overline{{\rm ln} \ V_L} \simeq -\rho_{typ} \ {\rm ln}L$, in agreement with previous numerical measures of $\alpha_{typ} =d+\rho_{typ} \simeq 4$ for the singularity spectrum $f(\alpha)$ of individual eigenfunctions. We also present numerical results concerning critical surface properties.

\end{abstract}

\maketitle

\section{ Introduction }

In statistical physics, any large-scale universal behavior is expected to 
come from some underlying renormalization ('RG') procedure that eliminates
all the details of microscopic models. In the presence of quenched 
disorder, interesting universal scaling behaviors usually occur both at phase
transitions (as in pure systems) but also in the low-temperature 
disorder-dominated phases. Since the main property of frozen disorder
is to break the translational invariance, the most natural renormalization
procedures that allow to describe spatial heterogeneities
are a priori real-space RG procedures \cite{realspaceRG}.
However, real-space RG 
such as the Migdal-Kadanoff block renormalizations \cite{MKRG},
 contain some approximations for most disordered models of interest
(these RG procedures become exact only for certain hierarchical lattices
\cite{berker,hierarchical}).
In this respect, an important exception is provided by
Anderson localization \cite{anderson}
which has remained a very active field
of research over the years
(see the reviews \cite{thouless,souillard,bookpastur,
Kramer,janssenrevue,markos,mirlinrevue}) :
for the usual Anderson tight binding model in arbitrary dimension $d$,
Aoki \cite{aoki80,aoki82,aokibook} has proposed
 an exact real-space renormalization (RG) procedure
at fixed energy
that preserves the Green functions for the remaining sites
(see more details in Section \ref{secrg} below).
However, the numerical results on the RG flows
obtained by Aoki thirty years ago were limited
to systems of linear sizes $L \leq 16$ in dimension $d=2$
 \cite{aoki80}, $L \leq 8$ in dimension $d=3$  \cite{aoki80,aoki82}
and to a very small statistics over the samples.
The aim of the present paper is thus to obtain more detailed
numerical results concerning
the statistics of renormalized on-site energies and renormalized 
hoppings for Anderson tight-binding model 
in dimension $d=2$, where only the localized phase exists,
and in dimension $d=3$, where there exists an Anderson transition.
Our main conclusions are the following :
(i) in the localized phase in dimension $d=2,3$, the statistics
of renormalized hoppings is not log-normal 
(in contrast with the conclusions of
 \cite{aoki80,aoki82} based on numerics on too small systems),
but involves the same universal properties
as the directed polymer model in dimension $1+(d-1)$,
in agreement with \cite{NSS,medina,prior}
(ii) at criticality, the statistics
of renormalized hoppings is multifractal in direct relation
with the multifractality of eigenstates (see the reviews
\cite{janssenrevue,mirlinrevue})
and the multifractality
of the two-point transmission 
\cite{janssen99,us_twopoints,us_manywires}.

The paper is organized as follows.
In Section \ref{secrg}, we describe the exact renormalization rules for Anderson models at fixed energy, and explain the physical meaning of renormalized observables
in terms of the Green function.
The statistical properties
of renormalized on-site energies is discussed in Section \ref{seceps}.
The statistics of renormalized hoppings 
is studied in the localized phase in Section \ref{secvloc},
and at criticality in Section \ref{secvcriti}.
Our conclusions are summarized in Section \ref{secconclusion}.

\section{ Real-space Renormalization rules at fixed energy }

\label{secrg}

\subsection{ Anderson localization Models }

The renormalization (RG) procedure described below
can be applied to any Anderson localization model of the generic form
\begin{eqnarray}
H = \sum_i \epsilon_i \vert i > < i \vert 
 +  \sum_{i,j}  V_{i,j} \vert i > < j \vert
\label{Hgene}
\end{eqnarray}
where $\epsilon_i$ is the on-site energy of site $i$
and where $V_{i,j}$ is the hopping between the sites $i$ and $j$.

\subsubsection{ Anderson tight binding model in dimension $d=2$ and $d=3$ }

\label{sectight}

The usual Anderson tight-binding model \cite{anderson} corresponds to 
the case where 

(a) the sites $(i)$ live on an hypercubic lattice in dimension $d$ 

(b) the hopping $V_{i,j}$ is unity if $i$ and $j$ are nearest neighbors
(and zero otherwise)

(c) the on-site energies $\epsilon_i$ are independent random variables
drawn from the flat distribution 
\begin{eqnarray}
p(\epsilon_i) = \frac{1}{W}
 \theta \left( - \frac{W}{2} \leq \epsilon_i \leq \frac{W}{2}  \right)
\label{flat}
\end{eqnarray}
The width $W$ thus represents the initial disorder strength.
It is known that in dimension $d=1,2$, only the localized phase exists,
whereas in dimension $d=3$, there exists an Anderson transition 
at some critical disorder $W_c$ whose numerical value is around
(see the review \cite{markos} and references therein)
\begin{eqnarray}
W_c \simeq 16.5
\label{wc3d}
\end{eqnarray}

\subsubsection{ Power-law Random Banded Matrix (PRBM) model}

\label{secpower}

The Power-law Random Banded Matrix (PRBM) model
is defined as follows : 
  the matrix elements $V_{i,j}$ are independent Gaussian
variables of zero-mean $\overline{V_{i,j}}=0$ and of variance
\begin{eqnarray}
\overline{ V_{i,j}^2 } = \frac{1}{1+ \left( \frac{r_{i,j}}{b}\right)^{2a}}
\label{defab}
\end{eqnarray}
where $r_{i,j}$ is the distance between sites $i$ and $j$.
One may consider either a line geometry with $r_{i,j}= \vert j-i \vert$
or the ring geometry of size $L$ (periodic boundary conditions) with
\begin{eqnarray}
r_{i,j}^{(L)} = \frac{L}{\pi} \sin \left( \frac{ \pi (i-j) }{L} \right)
\label{rijcyclic}
\end{eqnarray}
We refer to our recent works \cite{us_twopoints}, \cite{us_manywires}
for more details and references on the PRBM model.
The most important property is that the
 value of the exponent $a$ determines the localization properties
\cite{mirlin96} : 
 for $a>1$ states are localized with integrable power-law tails,
whereas for $a<1$ states are delocalized.
At criticality $a=1$, states become multifractal 
\cite{mirlin_evers,cuevas01,cuevas01bis,varga02}

\subsection{ RG rules upon the elimination of one site }

We now consider the Schr\"odinger equation
 at a given energy $E$ for an Hamiltonian of the form of Eq. \ref{Hgene}.
To eliminate a site $i_0$, we use may the 
Schr\"odinger equation projected on this site 
\begin{eqnarray}
E \psi(i_0) = \epsilon_{i_0} \psi(i_0) +\sum_j V_{i_0,j} \psi(j)
\label{psii0}
\end{eqnarray}
to make the substitution
\begin{eqnarray}
 \psi(i_0)=  \frac{1}{E-\epsilon_{i_0}} \sum_j V_{i_0,j} \psi(j)
\label{psii0elim}
\end{eqnarray}
in all other remaining equations.
Then from the point of view of other sites, any factor of the form
$V_{i,i_0} \psi(i_0)$ has to be replaced by 
\begin{eqnarray}
V_{i,i_0} \psi(i_0) \to  \frac{V_{i,i_0}}{E-\epsilon_{i_0}} \sum_j V_{i_0,j} \psi(j)
\label{rulevinter}
\end{eqnarray}
i.e. the hoppings between two neighbors $(i,j)$
of $i_0$ are renormalized according to
\begin{eqnarray}
V_{i,j}^{new} = V_{i,j} + \frac{V_{i,i_0} V_{i_0,j} }{E-\epsilon_{i_0}}
\label{rulev}
\end{eqnarray}
and the on-site energy of each neighbor $i$ of $i_0$ 
is renormalized according to
\begin{eqnarray}
\epsilon_{i}^{new} =\epsilon_{i} 
+ \frac{V_{i,i_0} V_{i_0,i} }{E-\epsilon_{i_0}}
\label{rulee}
\end{eqnarray}
These renormalizations equations are exact since they are based on
elimination of the variable $\psi(i_0)$ in the Schr\"odinger Equation.
The RG rules of Eqs \ref{rulev} and \ref{rulee} have been
introduced by Aoki \cite{aoki80,aoki82}
by considering the equations satisfied by the Green function.
Here we have chosen to derive them in the most elementary way
by direct substitution in the Schr\"odinger equation
to make obvious their origin and their exactness.

As stressed by Aoki \cite{aoki80,aoki82}, the RG rules of 
 Eqs \ref{rulev} and \ref{rulee} preserve the Green function
for the remaining sites.
This means for instance that if external leads
are attached to all surviving sites, the scattering properties
will be exactly determined using the renormalized parameters. 
To get a better intuition of the physical meaning of the renormalized
parameters, it is thus interesting to consider the simplest cases
where the disordered system is coupled to only one or two external wires
as we now describe.

\subsection{ Physical meaning of the renormalized on-site energies}

If one uses the RG rules of Eqs \ref{rulev} and \ref{rulee}
until there remains only a single site called $A$, the only
remaining parameter is the renormalized on-site energy $\epsilon_A(E)$.
If an external wire is attached to this site $A$,
the scattering eigenstate $\vert \psi >$  satisfies 
the  Schr\"odinger equation
\begin{eqnarray}
H \vert \psi > = E \vert \psi > 
\label{schodinger}
\end{eqnarray}  
inside the disorder sample and in the perfect wire
characterized by no on-site energy and by hopping unity between nearest
neighbors.
Within the wire, one has thus the plane-wave form
\begin{eqnarray}
\psi(x \leq x_{A}) && = e^{ik (x-x_A)} +r e^{- i k (x-x_A)}
\label{psiwirein}
\end{eqnarray}
where the energy $E$ is related to the wave vector $k$ by
\begin{eqnarray}
 E=2 \cos k  
\label{relationEk}
\end{eqnarray}
The reflexion coefficient $r$ of Eq. \ref{psiwirein}
is determined by the ratio
\begin{eqnarray}
R \equiv \frac{\psi(x_A-1)}{\psi(x_A)} = \frac{e^{-ik} + r e^{ik} }{1+r}
\label{ricc}
\end{eqnarray}
that is imposed by the Schr\"odinger Eq. \ref{schodinger}
projected onto site $A$. This can be computed in two ways as we now discuss.

\subsubsection{ Solution in terms of the renormalized on-site energy}

In terms of the renormalized on-site energy $\epsilon_A(E)$,
the Schr\"odinger Eq. \ref{schodinger}
projected onto site $A$ simply reads
\begin{eqnarray}
E  \psi(x_A)= \epsilon_A (E) \psi(x_A) +  \psi(x_A-1)
\label{eqzero}
\end{eqnarray}
i.e. the ratio of Eq. \ref{ricc} is directly related to the
renormalized on-site energy $\epsilon_A$
\begin{eqnarray}
R  = E- \epsilon_A(E)
\label{riccrenormeps}
\end{eqnarray}

\subsubsection{ Solution in terms of the spectrum of the closed system }

We denote by $(E_n,\phi_n)$ the spectrum of the disordered closed system,
so that the Hamiltonian inside the disordered sample reads
\begin{eqnarray}
H_{system} = \sum_n E_n \vert \phi_n ><\phi_n \vert
\label{closed}
\end{eqnarray}
In the presence of the wire, the scattering state
 $\vert \psi >$  of Eq. \ref{schodinger}
which takes the form of Eq. \ref{psiwirein} in the wire,
can be decomposed within the disordered system on the $(\phi_n)$ basis
 \begin{eqnarray}
\vert \psi> = \sum_n \alpha_n \vert \phi_n >
\label{dvpsi}
\end{eqnarray}
Projecting the Schr\"odinger Eq. \ref{schodinger} on $<\phi_m\vert$
yields the coefficients
\begin{eqnarray}
\alpha_m =  \frac{ \phi_m^*(x_A) \psi(x_A-1) } { E-E_m }
\end{eqnarray}
In particular at the contact point $A$, one obtains
\begin{eqnarray}
\psi(x_A)= \sum_n \alpha_n \phi_n (x_A) =
  \psi(x_A-1) \sum_n   \frac{ \vert \phi_n(x_A) \vert^2} { E-E_n }
\end{eqnarray}
so that the ratio $R$ of Eq. \ref{ricc} reads
\begin{eqnarray}
\frac{1}{R} =
 \sum_n   \frac{ \vert \phi_n(x_A) \vert^2} { E-E_n } \equiv G_E(x_A,x_A)
\label{riccinv}
\end{eqnarray}
in terms of the Green function $G_E$ of the closed system.

\subsubsection{ Relation between the on-site energy and the Green function}

In conclusion, the comparison of Eqs \ref{riccrenormeps} and \ref{riccinv}
yields
\begin{eqnarray}
\frac{1}{ E- \epsilon_A(E) } =  G_E(x_A,x_A) =  \sum_n   \frac{ \vert \phi_n(x_A) \vert^2} { E-E_n }
\label{epsrenormdv}
\end{eqnarray}
i.e. the on site-energy $\epsilon_A(E)$ of the remaining site $A$
is directly related to the Green function $G_E(x_A,x_A)$
at coinciding points.

\subsection{ Physical meaning of the renormalized hoppings}

If one uses the RG rules of Eqs \ref{rulev} and \ref{rulee}
until there remains only two sites called $A$ and $B$, the only
remaining parameters the two renormalized on-site energies $\epsilon_A(E)$,
$\epsilon_B(E)$ and the renormalized hoppings $V_{AB}(E)$.

\subsubsection{ Solution in terms of the renormalized parameters}

In terms of the renormalized parameters,
the Schr\"odinger Eq. \ref{schodinger}
projected onto sites $A$ and $B$ simply reads
\begin{eqnarray}
E  \psi(x_A)= \epsilon_A (E) \psi(x_A) +  \psi(x_A-1) +V_{AB}(E) \psi(x_B) 
\nonumber \\
E  \psi(x_B)= \epsilon_B (E) \psi(x_B) +  \psi(x_B+1) +V_{BA}(E) \psi(x_A) 
\label{schrAetB}
\end{eqnarray}
If two external wires are attached to $A$ and $B$
the scattering eigenstate $\vert \psi >$  satisfies 
the  Schr\"odinger Eq. \ref{schodinger}
inside the disorder sample and in the perfect wires,
characterized by no on-site energy and by hopping unity between nearest
neighbors, where one requires the plane-wave forms
\begin{eqnarray}
\psi_{in}(x \leq x_{A}) && = e^{ik (x-x_A)} +r e^{- i k (x-x_A)} \\
\nonumber \\
 \psi_{out}(x \geq x_B) && = t e^{ik (x-x_B)} 
\label{psiwires}
\end{eqnarray}
These boundary conditions define
 the reflection amplitude $r$ of the incoming wire
and the transmission amplitude $t$ of the outgoing wire.
The boundary conditions of Eq. \ref{psiwires}
determine the following ratio on the outgoing wire
\begin{eqnarray}
\frac{ \psi(x_{B}+1)}{\psi(x_{B}) } = e^{ik}
\label{outratio}
\end{eqnarray}
The following ratio 
\begin{eqnarray}
R \equiv \frac{\psi(x_{A}-1)}{ \psi(x_{A})}
\label{riccdef}
\end{eqnarray}
concerning the incoming wire can be then computed
in terms of the three real renormalized parameters from Eq. \ref{schrAetB}
\begin{eqnarray}
R = E - \epsilon_A
 - \frac{V_{AB}^2}{E-  \left( \epsilon_B + e^{ik} \right)}
\label{ricc2fils}
\end{eqnarray}
The reflexion coefficient $r$ of Eq. \ref{psiwires}
is then obtained as
\begin{eqnarray}
r = \frac{R-e^{-ik} }{e^{ik}-R}
\label{reflex}
\end{eqnarray}
yielding the Landauer transmission  
\begin{eqnarray}
T \equiv  \vert t \vert^2 = 1 - \vert r \vert^2
\label{deftotaltrans}
\end{eqnarray}

To simplify the discussion, we will focus in this paper on the case of
zero-energy $E=0$ (wave-vector $k=\pi/2$)
 that corresponds to the center of the band.
The Landauer transmission then reads in terms of the renormalized parameters
\begin{eqnarray}
T(E=0) 
= \frac{ 4 V_{AB}^2  (\epsilon_B^2+1)}
{ [\epsilon_A (\epsilon_B^2+1)-V_{AB}^2\epsilon_B]^2
+ [\epsilon_B^2+1+V_{AB}^2]^2}
\label{trg}
\end{eqnarray}

For later purposes, it is convenient to rewrite Eqs \ref{schrAetB}
as a system giving 
the values $\psi(x_A)$ and $\psi(x_B)$ at the contact points 
in terms of the values
$\psi(x_A-1)$ and $\psi(x_B+1)$ of the wires as
\begin{eqnarray}
\psi(x_A) && = \frac{1}{(E-\epsilon_A) D} \psi(x_A-1) 
+\frac{V_{AB}}{(E-\epsilon_A) (E-\epsilon_B) D} \psi(x_B+1) \\
\psi(x_B) && = \frac{V_{AB}}{(E-\epsilon_A) (E-\epsilon_B) D} \psi(x_A-1) 
+\frac{1}{(E-\epsilon_B) D}  \psi(x_B+1)
\label{sysout}
\end{eqnarray}
with the notation 
\begin{eqnarray}
D \equiv  1 - \frac{V_{AB}^2}{(E-\epsilon_A) (E-\epsilon_B) }
\label{deter}
\end{eqnarray}

\subsubsection{ Solution in terms of the spectrum of the closed system }

As above, we denote by $(E_n,\phi_n)$ 
the spectrum of the disordered closed system,
(Eq. \ref{closed}) and decompose the scattering state on the 
$(\phi_n)$ basis as in Eq. \ref{dvpsi}.
Projecting the Schr\"odinger Eq. \ref{schodinger} on $<\phi_m\vert$
yields the coefficients
\begin{eqnarray}
\alpha_m =  \frac{ \phi_m^*(x_A) \psi(x_A-1) } { E-E_m }
+  \frac{ \phi_m^*(x_B) \psi(x_B+1) } { E-E_m }
\end{eqnarray}
In particular at the contact points $A$ and $B$, one obtains
\begin{eqnarray}
\psi(x_A) && = \sum_n \alpha_n  \phi_n(A) = 
G_E(x_A,x_A)  \psi(x_A-1) + G_E(x_B,x_A)  \psi(x_B+1) \nonumber  \\
\psi(x_B) && = \sum_n \alpha_n  \phi_n(B) = G_E(x_A,x_B) \psi(x_A-1) + 
G_E(x_B,x_B)  \psi(x_B+1)
\label{system2f}
\end{eqnarray}
in terms of the Green function of the closed system
\begin{eqnarray}
G(i,j) =  \sum_{n \in L^d} \frac{\phi_n^*(i) \phi_n(j)}{E-E_n}
\label{defgreen}
\end{eqnarray}

\subsubsection{ Renormalized parameters in terms of the Green function}

In conclusion, the comparison between Eq. \ref{sysout} and \ref{system2f}
gives the Green functions in terms of the renormalized parameters
\begin{eqnarray}
G_{AA} && = \frac{1}{(E-\epsilon_A) D}  \nonumber \\
G_{BB} && = \frac{1}{(E-\epsilon_B) D} \nonumber \\
G_{AB} && = \frac{V_{AB}}{(E-\epsilon_A)(E- \epsilon_B) D} 
\label{Gidentif}
\end{eqnarray}
or by inversion the renormalized parameters in terms of the Green function
\begin{eqnarray}
E-\epsilon_A  && =  \frac{1}{ G_{AA} D}  \nonumber \\
E-\epsilon_B  && =  \frac{1}{ G_{BB} D}  \nonumber \\
V_{AB} && = \frac{G_{AB}}{ G_{AA} G_{BB} D} 
\label{espVidentif}
\end{eqnarray}
with
\begin{eqnarray}
D =  1 - \frac{V_{AB}^2}{(E-\epsilon_A)(E- \epsilon_B) }
= 1 - \frac{G_{AB}^2}{G_{AA} G_{BB}}
\label{deter2}
\end{eqnarray}
These relations clarify the physical meaning of the renormalized parameters
in terms of the Green functions that are usually considered
in the literature.

\subsection{ Numerical computations of renormalized parameters}

\begin{figure}[htbp]
 \includegraphics[height=6cm]{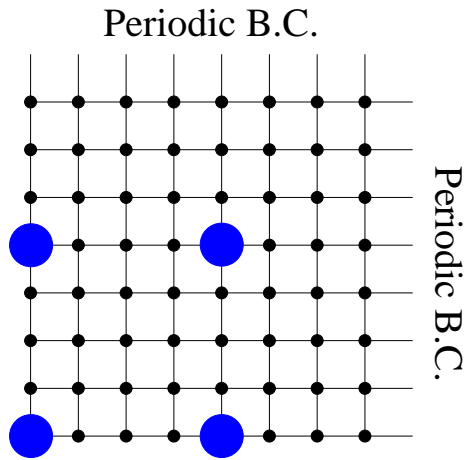}
\vspace{1cm}
\caption{ (Color on line) 
Renormalization procedure in dimension $d=2$.
The initial state is the tight-binding Anderson model
on a square lattice of size $L^2$,
 with periodic boundary conditions in the two directions.
Sites are then iteratively eliminated using 
 the RG rules of Eqs. 
\ref{rulev} and \ref{rulee} until there remains only the four sites
corresponding to the large discs, i.e. there are four renormalized on-site energies
and four renormalized hoppings at distance $L/2$ per sample. }
\label{figrg2d}
\end{figure}

\begin{figure}[htbp]
 \includegraphics[height=6cm]{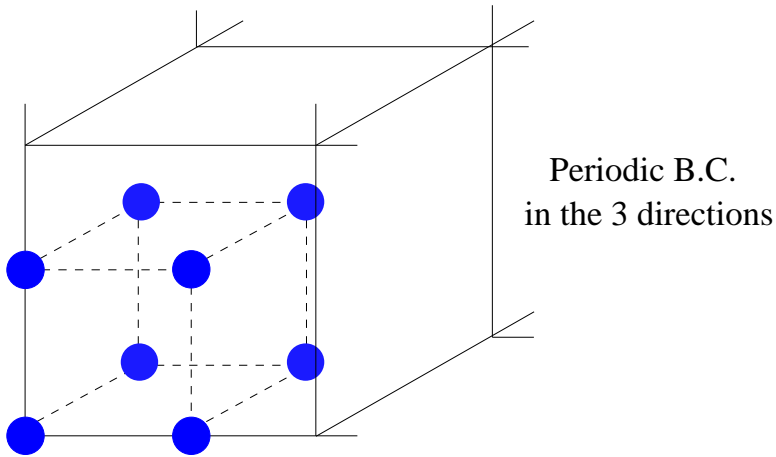}
\vspace{1cm}
\caption{ (Color on line) 
Renormalization procedure in dimension $d=3$.
The initial state is the tight-binding Anderson model
on a cubic lattice of size $L^3$,
 with periodic boundary conditions in the three directions
(here for clarity, the sites of the initial model have not be drawn
 in contrast to Fig. \ref{figrg2d} concerning the case $d=2$
which is more explicit).
Sites are then iteratively eliminated using 
 the RG rules of Eqs. 
\ref{rulev} and \ref{rulee} until there remains only the eight sites
corresponding to the large discs, i.e. there are eight
 renormalized on-site energies
and twelve renormalized hoppings at distance $L/2$ per sample. }
\label{figrg3d}
\end{figure}

\begin{figure}[htbp]
 \includegraphics[height=6cm]{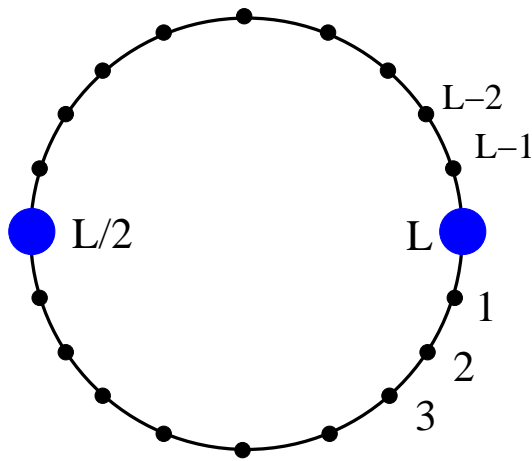}
\vspace{1cm}
\caption{ (Color on line) 
Renormalization procedure for the PRBM model with the ring geometry.
Sites are iteratively eliminated using the 
 the RG rules of Eqs. 
\ref{rulev} and \ref{rulee} until there remains only the two sites
$L/2$ and $L$
corresponding to the large discs, i.e. there are two
 renormalized on-site energies
and one renormalized hoppings per sample. }
\label{figrgring}
\end{figure}

The RG rules of Eqs \ref{rulev} and \ref{rulee}
 can be followed numerically from the initial condition
given by the model of Eq. \ref{Hgene} under interest.
 In the following, we describe the sizes
and the statistics over the samples that we have studied for the
 Anderson tight binding model in dimension $d=2$ and $d=3$
and the PRBM model.

\subsubsection{ Anderson tight binding model in dimension $d=2$ and $d=3$ }

\label{numericaldetails}

For the Anderson tight-binding model described in Section \ref{sectight},
we have followed numerically the RG rules 
starting from an hypercubic lattice of size $L^d$ with periodic
boundary conditions in all $d$ directions.
In each sample, the final state that we analyse is an 
hypercube of linear size $L/2$, as shown on Figure \ref{figrg2d}
for $d=2$ and on Figure \ref{figrg3d} for $d=3$

(i) in dimension $d=2$, there are four remaining sites per sample
as shown on Fig. \ref{figrg2d}, i.e. there are 
four renormalized on-site energies and four renormalized couplings
at distance $L/2$. 
We have studied the sizes $L =12,24,36,48,60,72,84,96,108,120$.
The corresponding numbers $ns(L)$ of independent
samples are
 of order $n_s(L=12) =2.10^7$, $n_s(L=60) =33.10^3$, $n_s(L=120) =1150$.

(ii) in $d=3$ there are eight remaining sites per sample
as shown on Fig. \ref{figrg3d}, i.e. there are
eight renormalized on-site energies and twelve renormalized couplings
at distance $L/2$.
We have studied the sizes $L=4,6,8,10,12,14,16,18,20,22,24,26,28,30$. 
 The corresponding numbers $n_s(L)$ of independent
samples are of order $n_s(L=4) =10^7$, $n_s(L=10) =6.10^4$,
  $n_s(L=20) =400$ and $n_s(L=30) =24$.

\subsubsection{ Power-law Random Banded Matrix (PRBM) model}

For the PRBM model described in Section \ref{secpower},
we have followed numerically the RG rules
up to the final state shown on Fig. \ref{figrgring}
containing only the sites $L/2$ and $L$, i.e. in each sample,
there are two renormalized on-site energies and one renormalized coupling.
We have studied rings of sizes $50 \leq L \leq 1800$
with corresponding statistics of $10.10^8 \geq n_s(L) \geq 2400$
 independent samples.

\section{ Statistics of renormalized on-site energies  }

\label{seceps}

\subsection{ General properties }

We find that the renormalized on-site energies remain finite in all phases
(localized, delocalized, critical), and that 
the histograms ${\cal P}_L(\epsilon)$ corresponding to various system sizes $L$
converge towards 
some stationary distribution ${\cal P}_{\infty}(\epsilon)$
that present the following common properties :

(i) ${\cal P}_{\infty}(\epsilon)$ is symmetric in $\epsilon \to - \epsilon$
(as the initial condition of Eq. \ref{flat})

(ii)  ${\cal P}_{\infty}(\epsilon)$ has a finite density ${\cal P}_{\infty}(0)$
at its center $\epsilon=0$. 
After the change of variables to $y \equiv \ln \vert \epsilon \vert$, 
this corresponds to 
\begin{eqnarray}
   P_{\infty}(y \equiv \ln \vert \epsilon \vert ) \oppropto_{y \to -\infty} 
\vert \epsilon \vert = e^{y}
\label{puleft}
\end{eqnarray}

(iii) For $\epsilon \to \pm \infty$, ${\cal P}_{\infty}(\epsilon)$
presents the following power-law decay  
\begin{eqnarray}
   {\cal P}_{\infty}(\epsilon ) \oppropto_{\epsilon \to \pm\infty} 
\frac{1}{\epsilon^2}
\label{pepsinfty}
\end{eqnarray}

After the change of variables to $y \equiv \ln \vert \epsilon \vert$,
Eq. \ref{pepsinfty} corresponds to 
\begin{eqnarray}
   P_{\infty}(y \equiv \ln \vert \epsilon \vert ) \oppropto_{y \to +\infty} 
\frac{1}{\vert \epsilon \vert} = e^{-y}
\label{puright}
\end{eqnarray}

The origin of the power-law of Eq. \ref{pepsinfty}, even when one starts
from a bounded distribution in $\epsilon$ as in the tight-binding
Anderson model (see Eq. \ref{flat}), can be understood from
the form the RG rule of Eq. \ref{rulee} which reads at zero energy $E=0$
\begin{eqnarray}
\epsilon_{i}^{new} =\epsilon_{i} 
- \frac{V_{i,i_0} V_{i_0,i} }{\epsilon_{i_0}}
\label{ruleezero}
\end{eqnarray}
During the first steps of renormalization where the hoppings $V$ are finite,
very large renormalized on-site energies are generated when 
the eliminated on-site energy $\epsilon_{i_0}$ is very small.
The finite density of ${\cal P}(\epsilon_{i_0})$ at $\epsilon_{i_0}=0$
yields the power-law decay of Eq. \ref{pepsinfty}
via the change of variable $\epsilon_{i}^{new} \simeq - 1/\epsilon_{i_0}$ 
using the standard formula $P_{new}(\epsilon_{i}^{new}) d \epsilon_{i}^{new}
= {\cal P}(\epsilon_{i_0}) d \epsilon_{i_0}$.

In the remaining of this section, we present the histograms we have
measured in various cases.

\subsection{ Results for the square lattice in dimension $d=2$ }

\begin{figure}[htbp]
 \includegraphics[height=6cm]{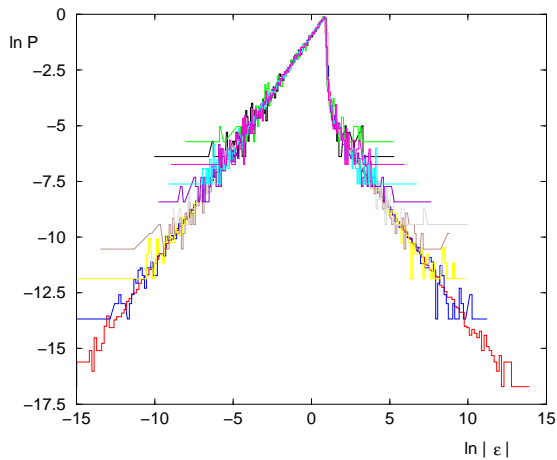}
\vspace{1cm}
\caption{ (Color on line) 
Statistics of renormalized on-site energies $\epsilon$
 for the Anderson model on the square lattice
in dimension $d=2$ for sizes $12 \leq L \leq 120$ (disorder strength $W=40$)
: the histograms of $\ln \vert \epsilon \vert$
are identical (apart for the cut-off imposed by the statistics
over the samples)
 The left and right slopes of value
unity corresponds to Eqs \ref{puleft} and \ref{puright}. }
\label{figcubic2deps}
\end{figure}

On Fig. \ref{figcubic2deps}, we show the histograms of the logarithm
of the absolute value of the renormalized on-site energy  $\epsilon$
for various sizes $12 \leq L \leq 120$ : apart from the cut-offs in the tails
imposed by different statistics over the samples, these histograms coincide.
This shows that the convergence towards the stationary distribution
 ${\cal P}_{\infty}(\epsilon )$ is quite rapid : starting from
the initial condition of Eq. \ref{flat}, our results for the smallest size
$l=12$ have already 'converged' towards the final- and very different- distribution of Figure \ref{figcubic2deps}. On Fig. \ref{figcubic2deps},
the slope of the left tail is of order $+1$ in agreement with Eq. \ref{puleft},
and the slope of the right tail is of order $-1$ in agreement with Eq. \ref{puright}.

\begin{figure}[htbp]
 \includegraphics[height=6cm]{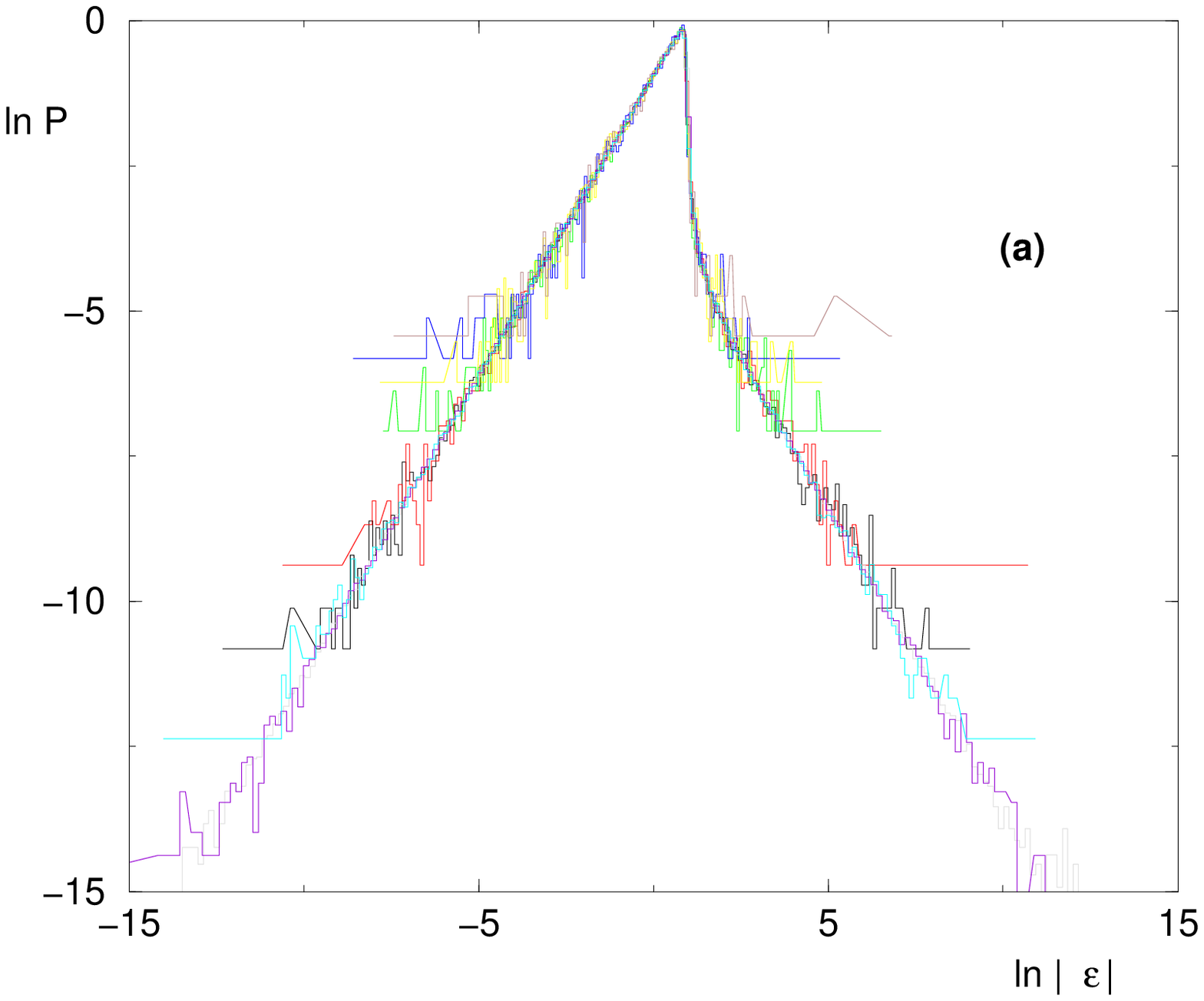}
\hspace{2cm}
 \includegraphics[height=6cm]{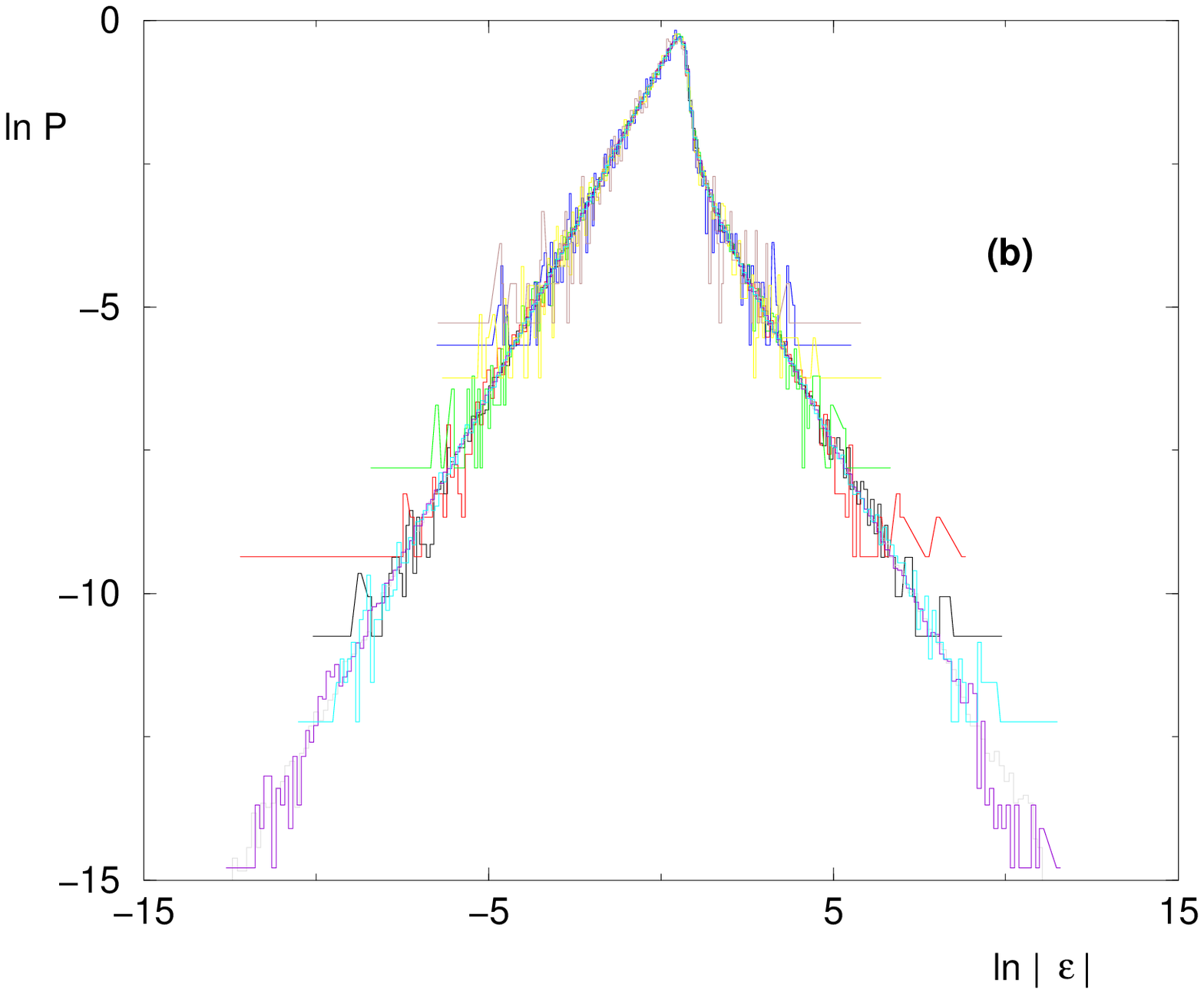}
\vspace{1cm}
\caption{ (Color on line) 
Statistics of renormalized on-site energies $\epsilon$
 for the Anderson model on the cubic lattice
in dimension $d=3$ for sizes $4 \leq L \leq 20$
(a) Histograms of $\ln \vert \epsilon \vert$ in the localized phase ($W=40$)
(b) Histograms of $\ln \vert \epsilon \vert$ at criticality ($W_c=16.5$) }
\label{fig3deps}
\end{figure}

\subsection{ Results for the cubic lattice in dimension $d=3$ }

Our data for the Anderson tight binding model in $d=3$ 
are shown on Fig. \ref{fig3deps} : 
both in the localized phase and at criticality,
the convergence in $L$ towards the stationary distribution 
${\cal P}_{\infty}(\epsilon )$ is still rapid, and the measured tails are 
again in agreement with Eqs \ref{puleft} and \ref{puright}.
It turns out that for a given disorder value,
our numerical results concerning ${\cal P}_{\infty}(\epsilon )$
seem to coincide for $d=2$ and $d=3$ (Fig. \ref{figcubic2deps}
and Fig. \ref{fig3deps} (a) corresponding to $W_d=40$) :
 the reasons of this coincidence are not clear to us,
since the initial coordinence of sites clearly depends on the dimension $d$.

\subsection{ Results for the PRBM model }

\begin{figure}[htbp]
 \includegraphics[height=6cm]{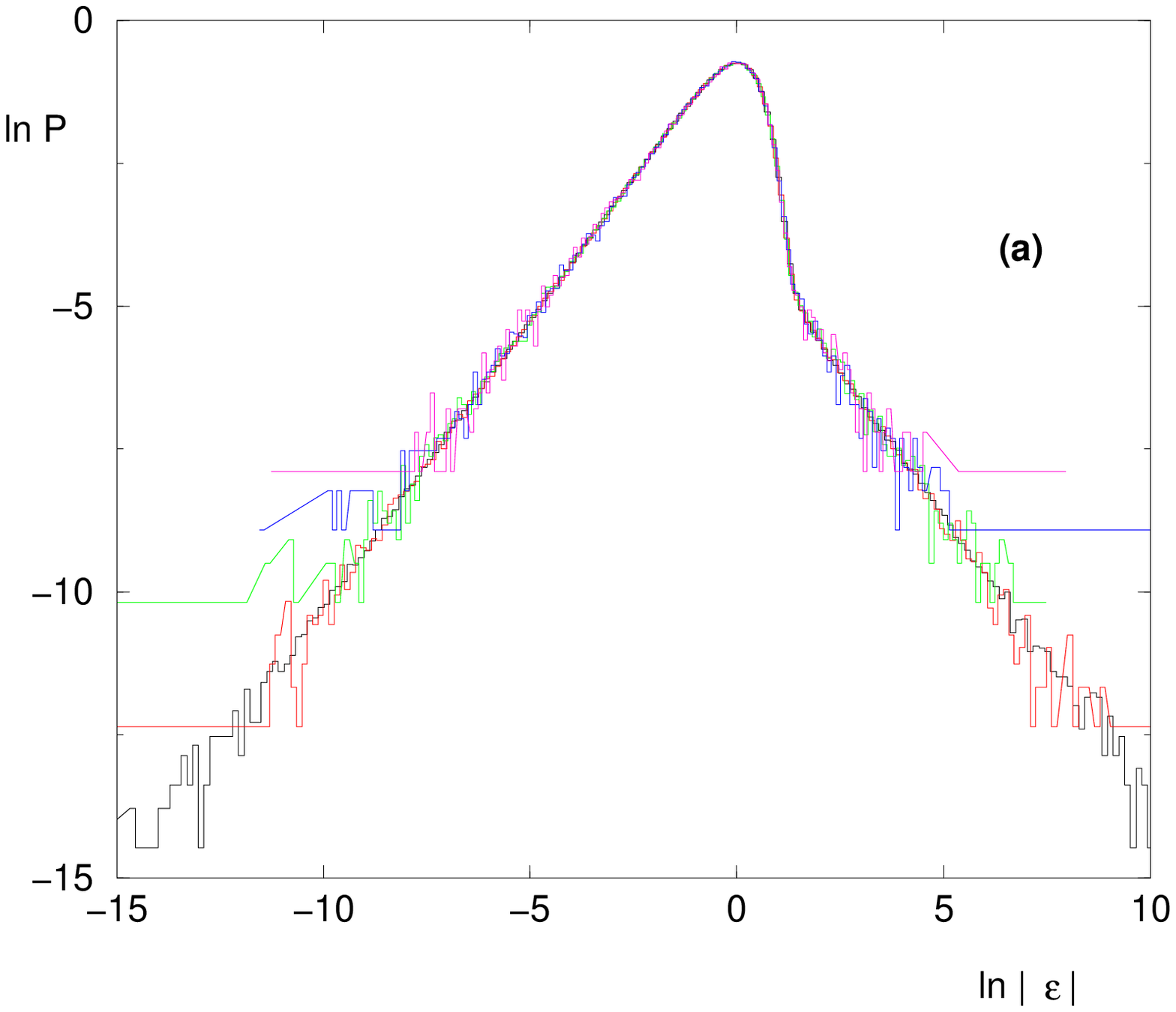}
\hspace{2cm}
 \includegraphics[height=6cm]{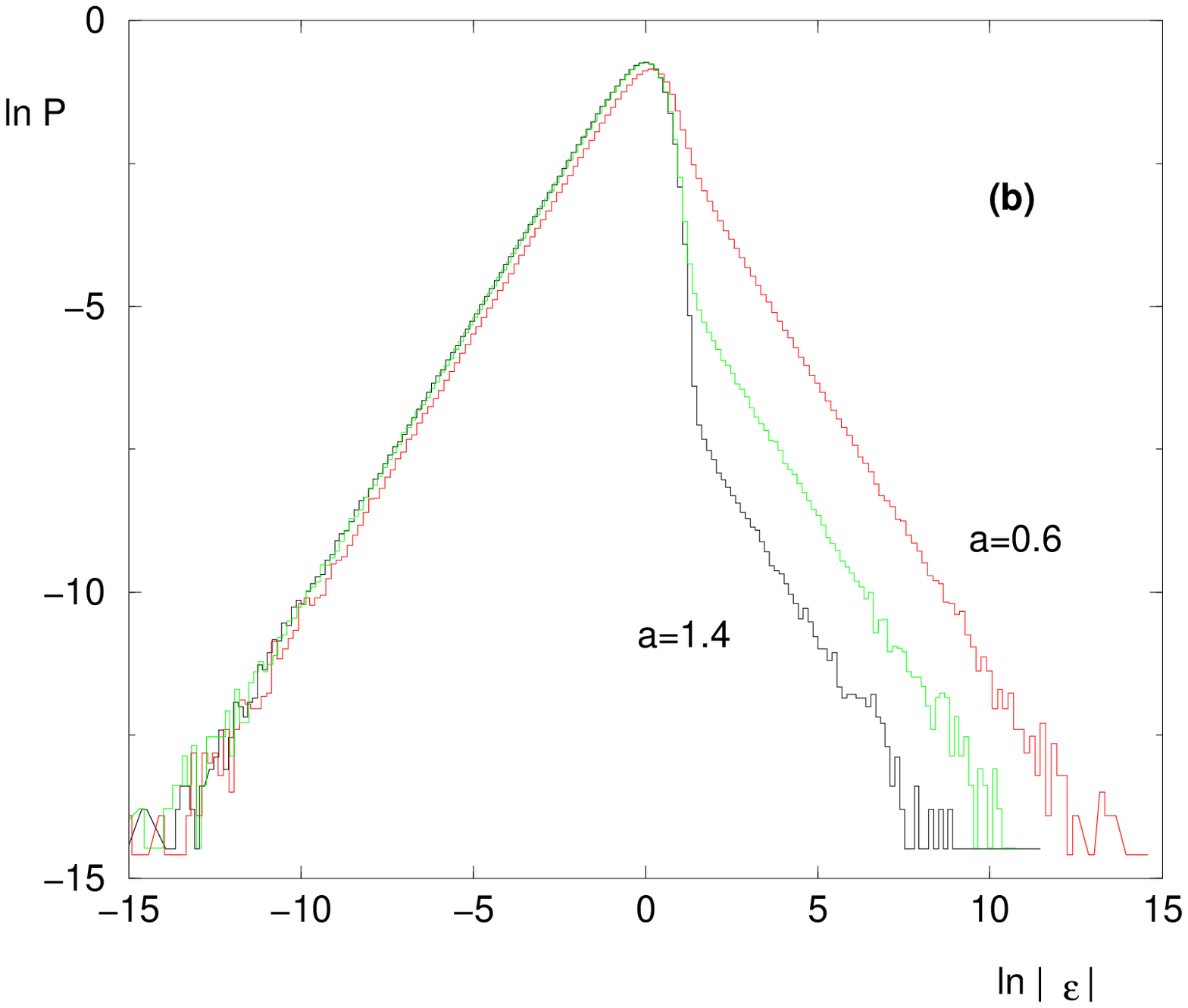}
\vspace{1cm}
\caption{ (Color on line) 
Statistics of renormalized on-site energies $\epsilon$
 in the PRBM model of parameter $b=0.1$
(a) The histograms of $\ln \vert \epsilon \vert$ 
at criticality $a=1$ for various sizes $L=100,200,400,600,800$ 
are identical (apart for the cut-off imposed by the statistics
over the samples).
(b) Comparison of the stationary distributions
 in the localized phase ($a=1.4$), at criticality  ($a=1$)
and in the delocalized phase ($a=0.6$). The left and right slopes of value
unity corresponds to Eqs \ref{puleft} and \ref{puright}. }
\label{figpowereps}
\end{figure}

The properties found above for Anderson tight-binding models
seem to be valid for more general
 Anderson models of the form of Eq. \ref{Hgene}.
As an example, we show on Fig. \ref{figpowereps} our data concerning
the PRBM model described in section \ref{secpower}.
The histograms of renormalized energies converge rapidly towards their limit.
The stationary distribution presents the tails of Eqs 
\ref{puleft} and \ref{puright} in all phases
(localized, critical, delocalized).

\subsection{Consequences }

In conclusion, the renormalized on-site energies remain finite random
variables in all phases (localized, critical, delocalized).
As a consequence, the behavior of the two-point Landauer transmission
of Eq. \ref{trg} is determined by the properties of the 
renormalized hoppings 

(i) in the delocalized phase, both the renormalized hopping
and the two-point transmission will remain random finite variables.

(ii) in the localized phase and at criticality
where the two-point transmission decays with the distance,
its decay
will be directly related to the decay of the renormalized hopping via
\begin{eqnarray}
\ln T(E=0) \simeq \ln  V_{AB}^2 + {\rm finite}
\label{trgdecay}
\end{eqnarray}

In the following, we discuss the statistics of renormalized hoppings
in the localized phase and at criticality, in relation with the statistics
of two-point transmission.

\section{ Statistics of renormalized hoppings in the localized phase }

\label{secvloc}

\subsection{ Universality class of the directed polymer in a random medium}

In dimension $d=1$, 
the transfer matrix formulation of the Schr\"odinger 
equation yields a log-normal distribution 
for the Landauer transmission \cite{anderson_fisher,luck}
\begin{eqnarray}
\ln T_L^{(d=1)} \oppropto_{L \to \infty} - \frac{L}{\xi_{loc}} + L^{1/2} u
\label{trans1d}
\end{eqnarray}
The leading non-random term is extensive in $L$ and involves
the localization length $\xi_{loc} $.
The subleading random term is of order $L^{1/2}$,
and the random variable $u$ of order $O(1)$ is Gaussian 
distributed as a consequence of the Central
Limit theorem.
Although it has been very often assumed and written
that this log-normal distribution 
persists in the localized phase in dimension $d=2,3$, 
theoretical arguments  \cite{NSS,medina}
and recent numerical calculations \cite{prior}
are in favor of the following scaling
 form for the logarithm of the transmission
\begin{eqnarray}
\ln T_L^{(d)} \oppropto_{L \to \infty} -  \frac{L}{\xi_{loc}} + L^{\omega(d)} u
\label{transd}
\end{eqnarray}
where the exponent $\omega(d)$ depends on the dimension $d$ 
and coincides with the droplet exponent 
characterizing the strong disorder
phase of the directed polymer in a
random medium of dimension $1+(d-1)$ (see the review \cite{Hal_Zha} on
directed polymers). The probability distribution of the rescaled variable $u$
is not Gaussian but is determined by the directed polymer
universality class (see \cite{prior} where its distribution in $d=2$
is shown to coincide with the exactly known Tracy-Widom distribution
for the directed polymer in $1+1$).

The arguments in favor of the same universality class
can be decomposed in two steps \cite{NSS,medina,prior} :

(i) in the localized phase of Anderson localization in dimension $d$,
the transmission decays exponentially with the length, and thus 
{\it directed paths } completely dominate asymptotically over non-directed
paths. In dimension $d=2$, the dominance of a narrow channel can be seen
on Figs 10 and 11 of Ref \cite{markostraj}.

(ii) these directed paths of the Anderson model
have weights that are random both in magnitude and sign,
but it turns out that the directed polymer model which is usually defined
with random positive weights (Boltzmann weights) keeps the same 
exponents in the presence of complex weights  
(see section 6.3 of the review \cite{Hal_Zha}).

In conclusion, from the relation of Eq. \ref{trgdecay}, we expect that the
renormalized hoppings will present the same statistics as 
the Landauer transmission of Eq. \ref{transd}
\begin{eqnarray}
 \ln V_L  \simeq - \frac{L}{\xi_{loc}} +  L^{\omega(d)} u +...
\label{vfluctloc}
\end{eqnarray}
To check this relation, we have measured
 the averaged value and the variance of 
the logarithm of the renormalized hoppings in dimension $d=2,3$.

\subsection{ Results for the square lattice in dimension $d=2$ }

\begin{figure}[htbp]
 \includegraphics[height=6cm]{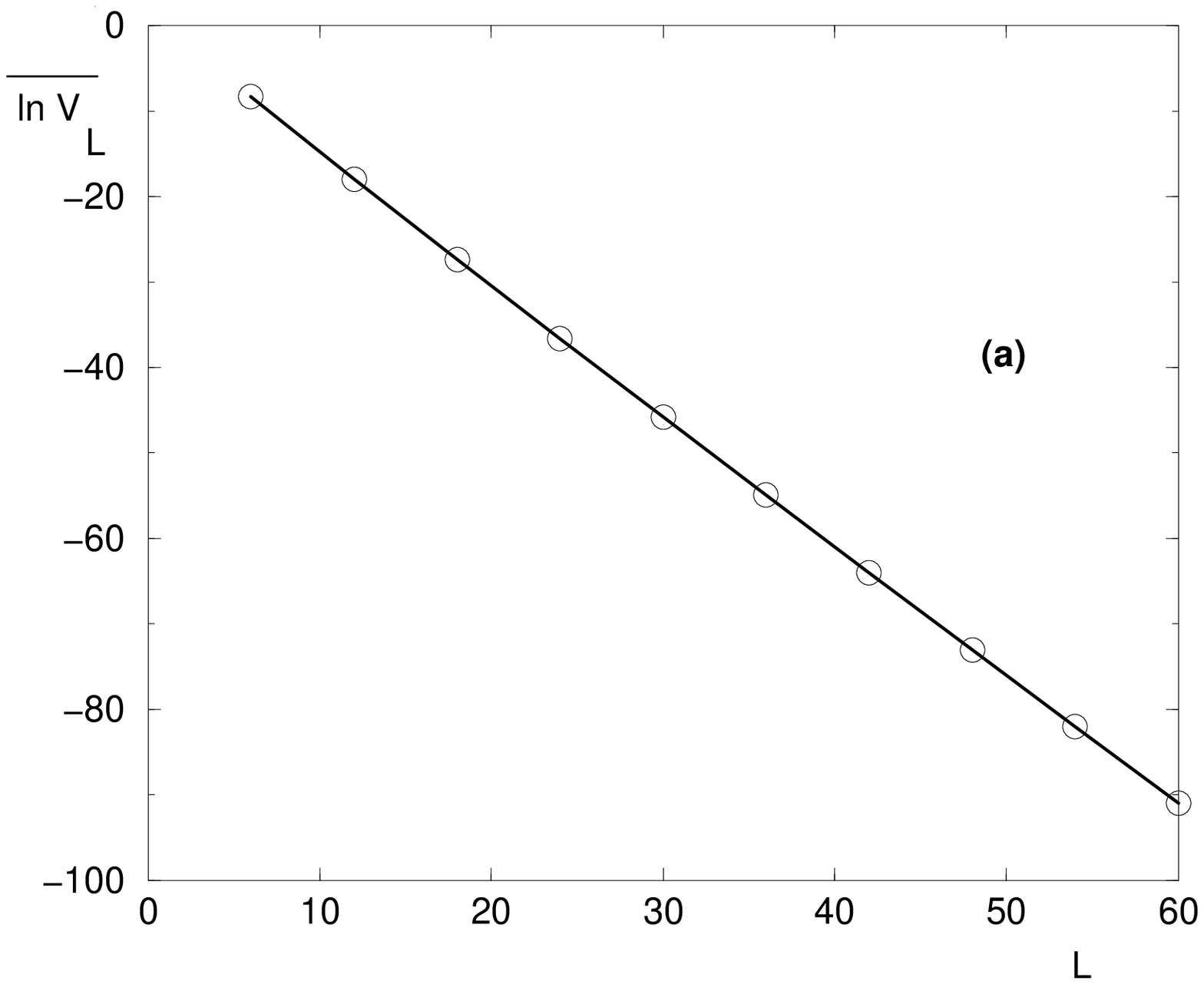}
\hspace{2cm}
 \includegraphics[height=6cm]{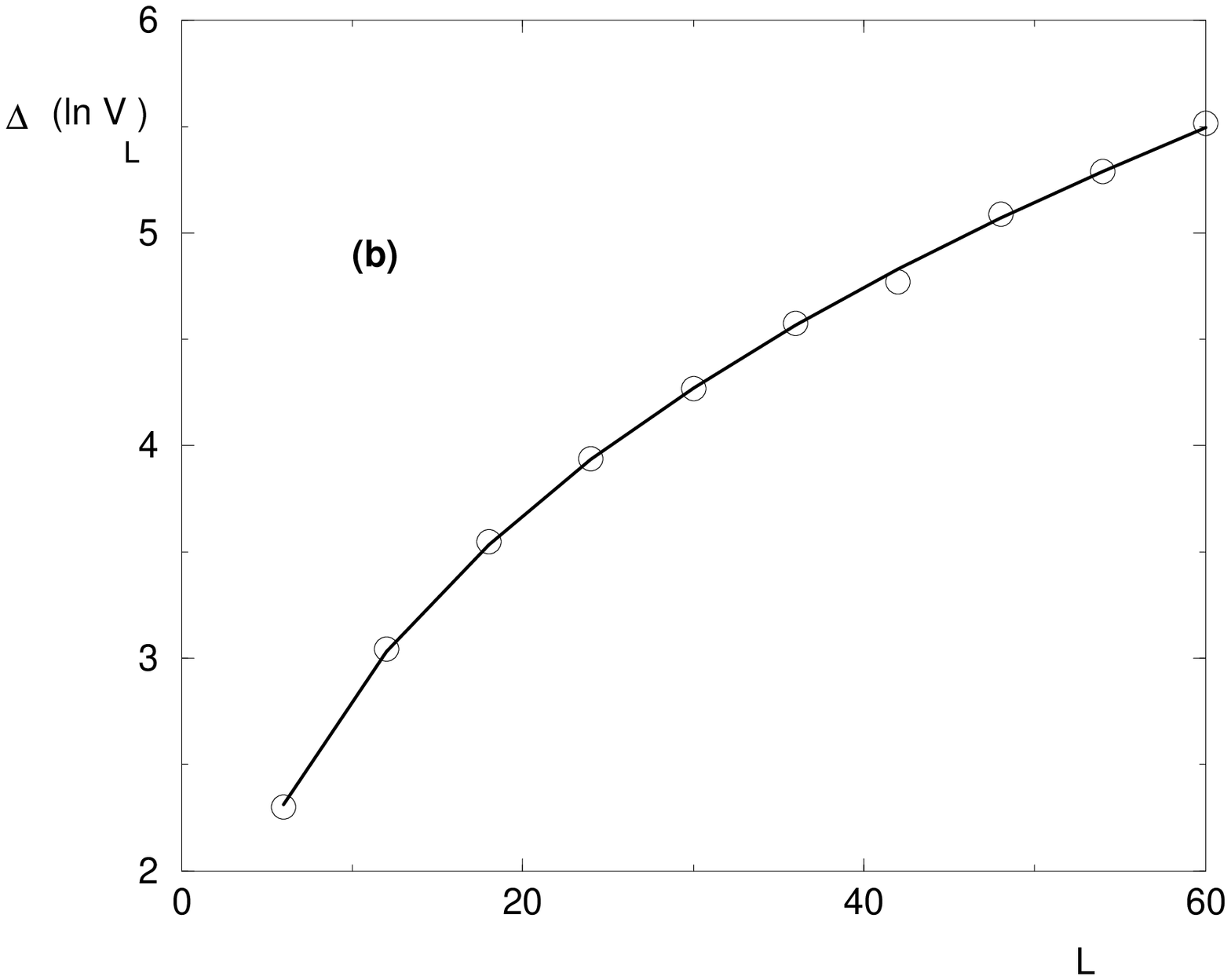}
\vspace{1cm}
\caption{
Statistics of the typical renormalized hopping
 in $d=2$ where only the localized phase exists
(the data shown correspond to the disorder strength $W=40$)
(a) Typical exponential decay :
$\overline{ \ln V_L}$ is linear in $L$ and
the slope represents the inverse of the localization length $\xi_{loc}$
(Eq. \ref{vfluctloc}).
(b) The fluctuation term
$\Delta (\ln V_L) $ grows as $L^{\omega}$
(Eq. \ref{vfluctloc}) with $\omega(d=2) \simeq 0.33$. }
\label{figvcubic2d}
\end{figure}

\begin{figure}[htbp]
 \includegraphics[height=6cm]{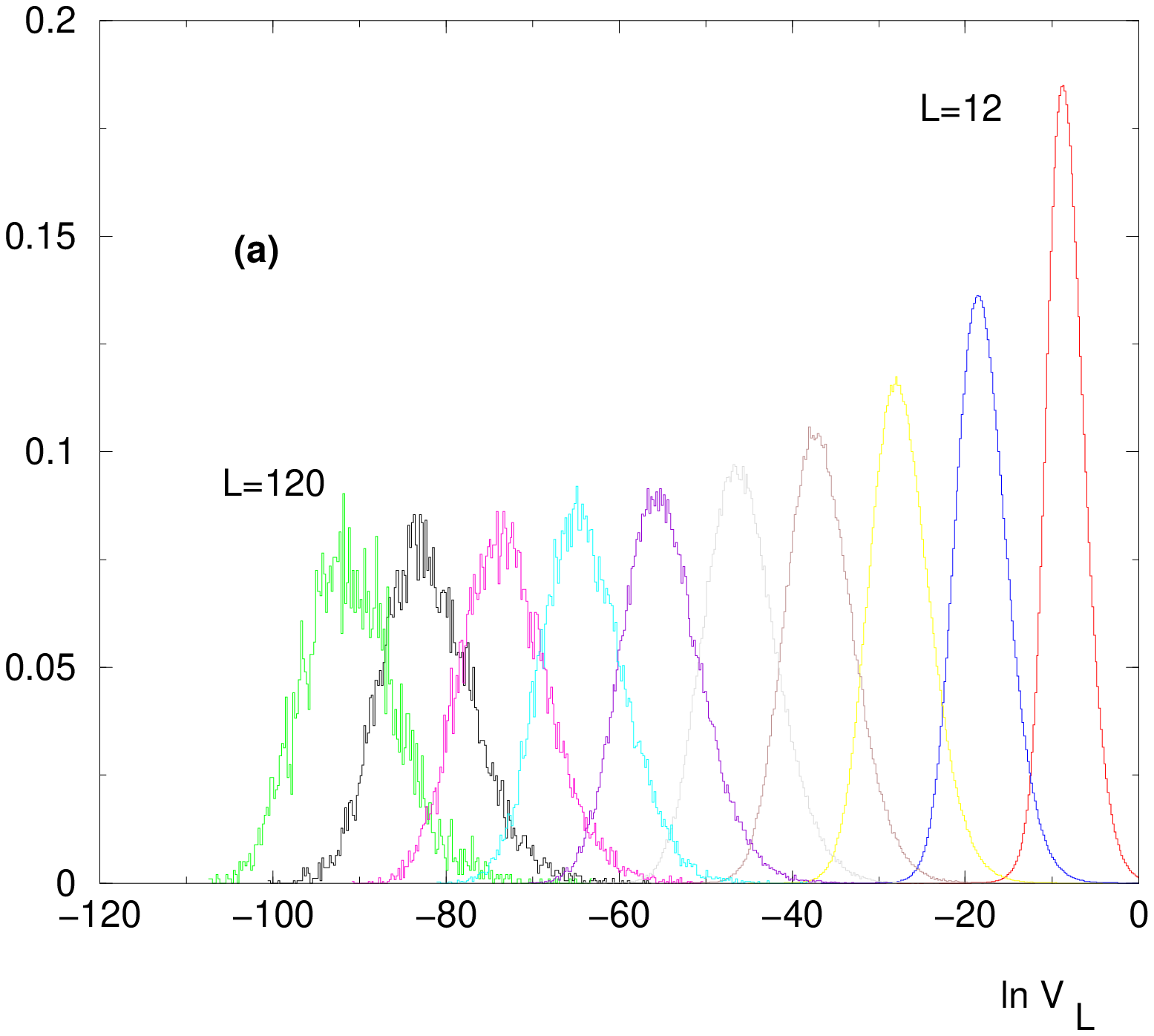}
\hspace{2cm}
 \includegraphics[height=6cm]{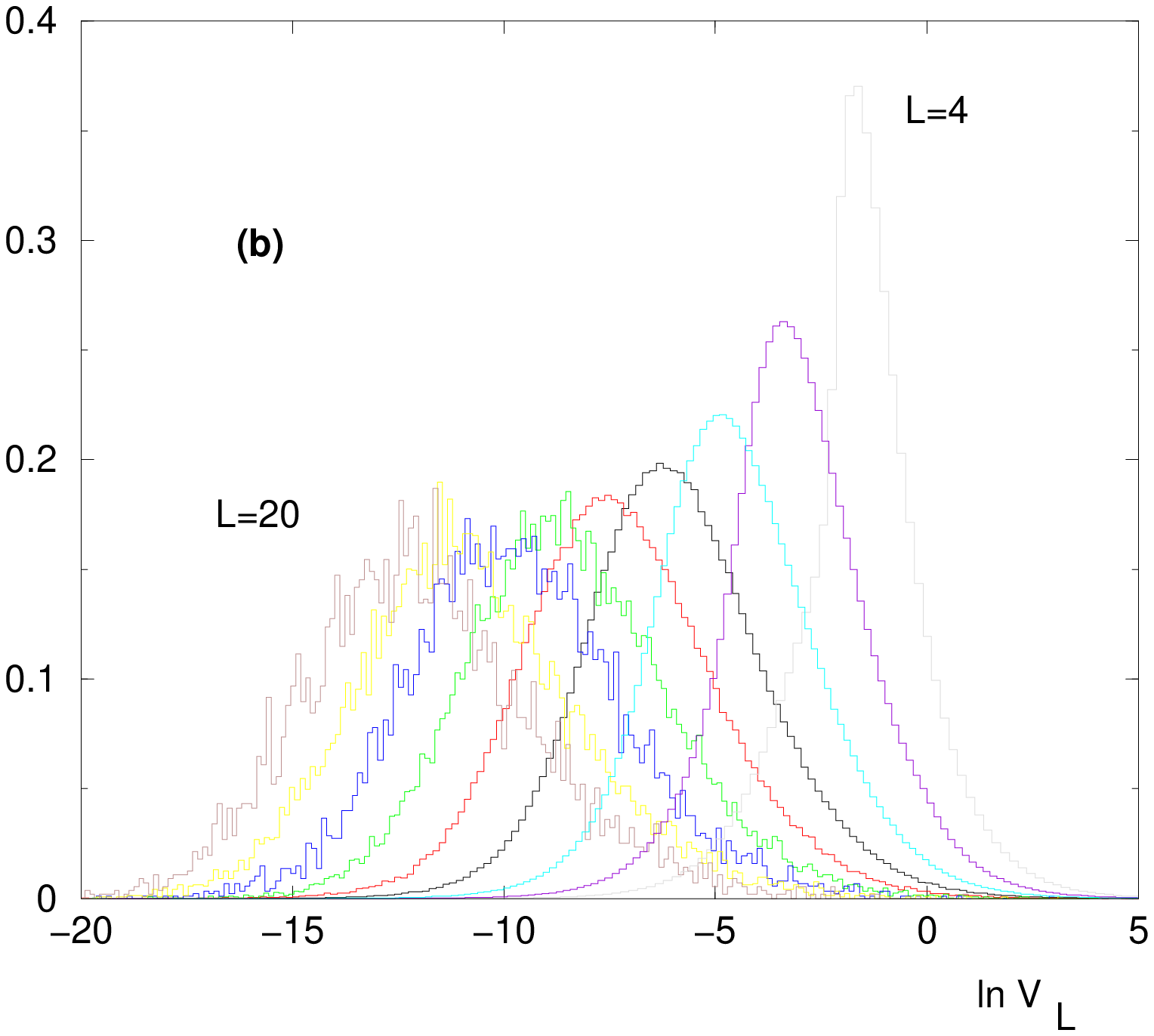}
\vspace{1cm}
\caption{ (Color on line) 
Histograms of the logarithm of the renormalized hopping $\ln V_L$
for various lengths $L$ in the localized phase 
(for the disorder strength $W=40$ 
(a)  in dimension $d=2$ 
(b)  in dimension $d=3$  }
\label{fighistocubic2d3d}
\end{figure}

In dimension $d=2$, only the localized phase exists.
On Fig. \ref{figvcubic2d} (a), we show the typical exponential decay
corresponding to a finite localization length $\xi_{loc}$ in Eq. \ref{vfluctloc}.
On Fig. \ref{figvcubic2d} (b), we show the amplitude $\Delta (\ln V_L) $
of the random term in Eq. \ref{vfluctloc} : the three parameters fit
$\Delta (\ln V_L) = a_0 L^{\omega(d=2)}+a_1$ yields the value
\begin{eqnarray}
 \omega(d=2) \simeq 0.33
\label{omegad2}
\end{eqnarray}
in agreement with the exact result \cite{Hus_Hen_Fis,Kar,Joh,Pra_Spo}
\begin{eqnarray}
 \omega_{DP}(1+1) =\frac{1}{3}
\label{omegadp2}
\end{eqnarray}
for the directed polymer in a random medium of dimension $1+1$.
On Fig. \ref{fighistocubic2d3d} (a), 
we show the histograms of $(\ln V_L)$ for various
sizes $L$ : as $L$ grows, the maximum moves linearly while the width grows as
$L^{\omega}$.

\subsection{ Results for the cubic lattice in dimension $d=3$ }

\begin{figure}[htbp]
 \includegraphics[height=6cm]{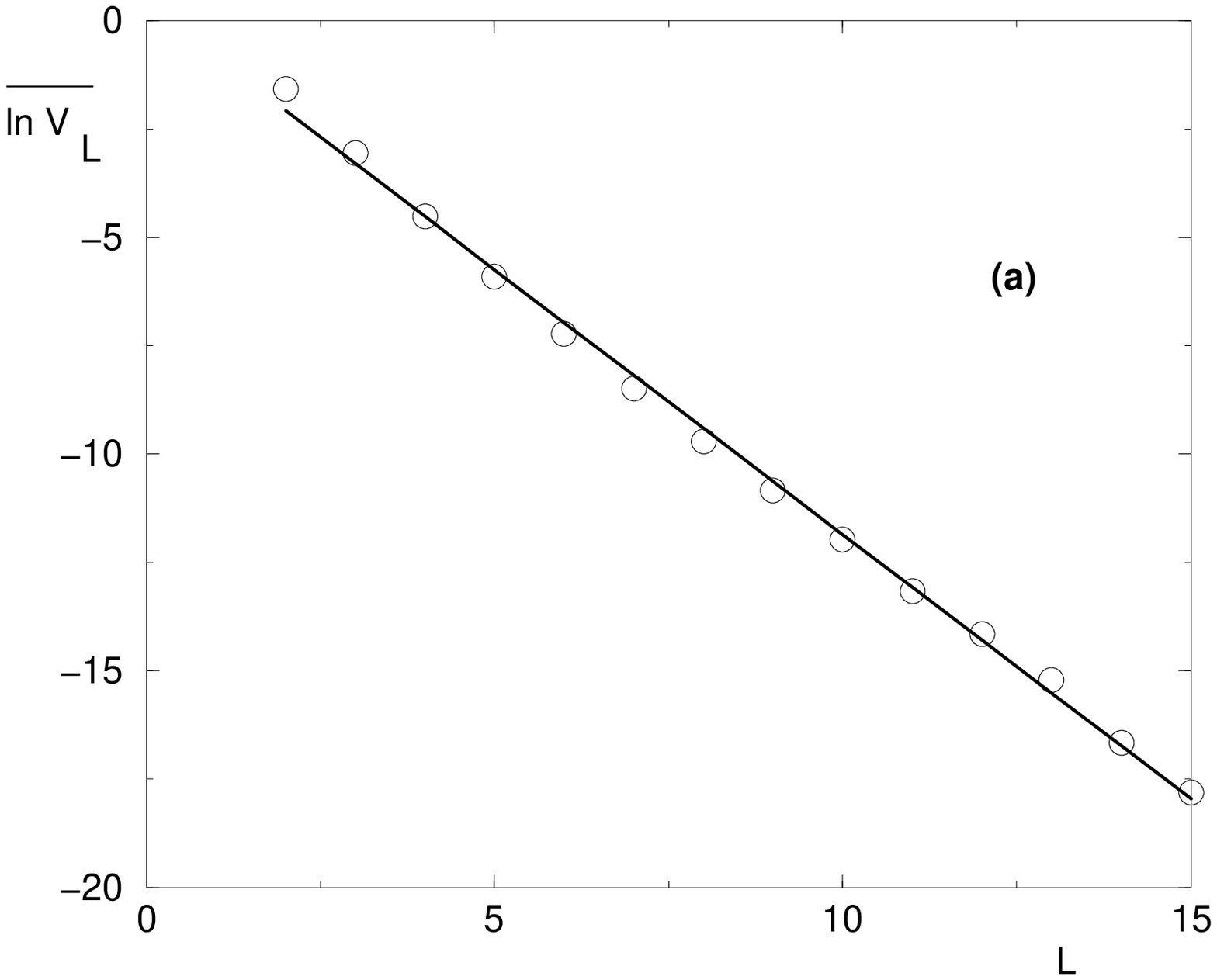}
\hspace{2cm}
 \includegraphics[height=6cm]{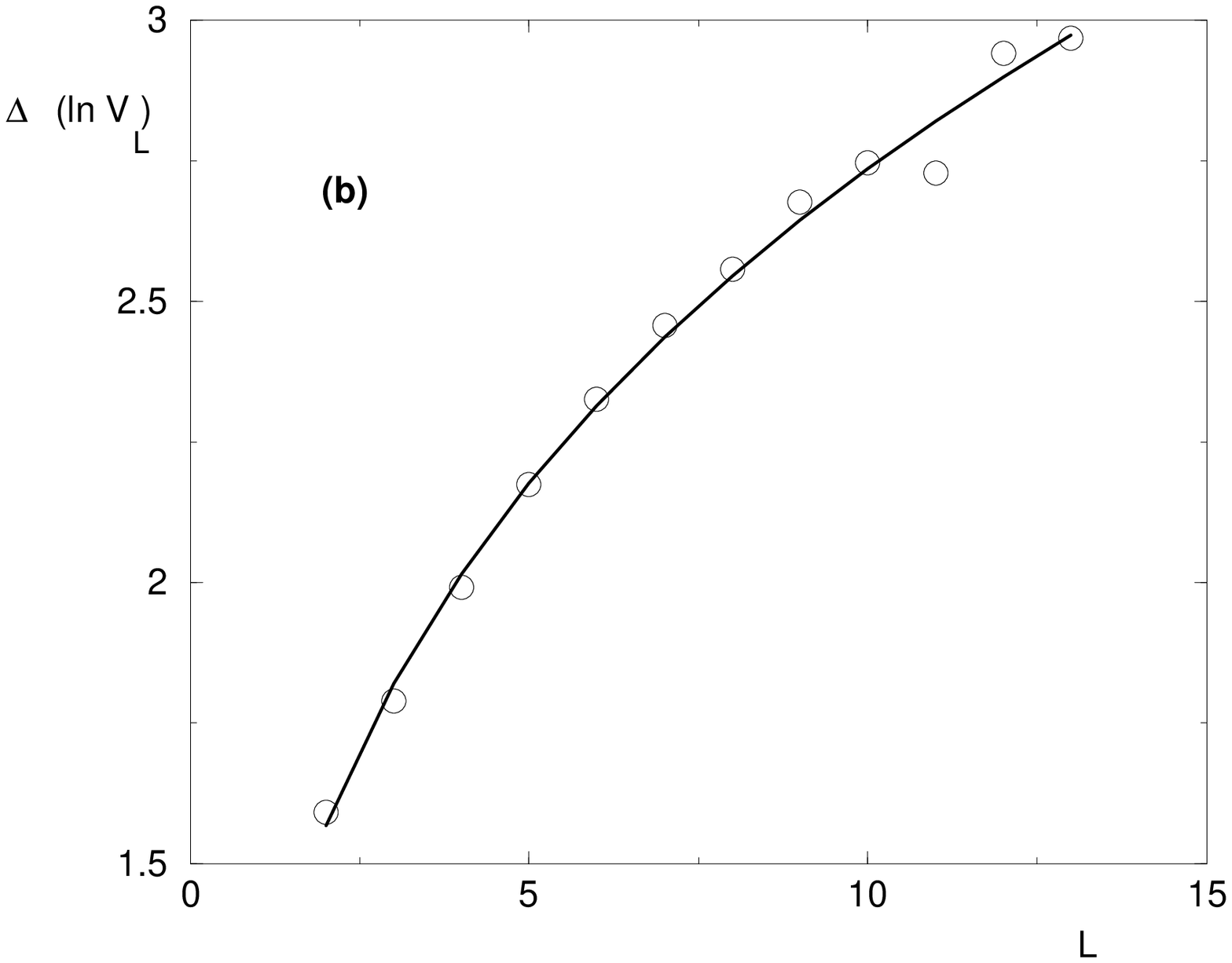}
\vspace{1cm}
\caption{ 
Statistics of the typical renormalized hopping
 in $d=3$ in the localized phase ($W=40$)
(a) $\overline{ \ln V_L}$ as a function of $L$ :
the slope represents the inverse of the localization length $\xi_{loc}$
(Eq. \ref{vfluctloc})
(b)  The fluctuation term
$\Delta (\ln V_L)  $ grows as $L^{\omega}$
(Eq. \ref{vfluctloc}) with $\omega(d=3) \simeq 0.24$.}
\label{figvcubic3dloc}
\end{figure}

In dimension $d=3$, the localized phase exists in the domain 
$W>W_c \simeq 16.5$ for the disorder strength (Eq. \ref{wc3d}).
The data shown on Fig. \ref{figvcubic3dloc} corresponds to the
disorder strength $W=40$.
The histograms of $(\ln V_L)$ for various sizes $L$ are shown on Fig. 
 \ref{fighistocubic2d3d} (b).
On Fig. \ref{figvcubic3dloc} (a), the typical exponential decay
found corresponds to a finite localization length $\xi_{loc}$ in Eq. \ref{vfluctloc}.
On Fig. \ref{figvcubic3dloc} (b), the amplitude $\Delta (\ln V_L) $
of the random term in Eq. \ref{vfluctloc} can be fitted
by the form 
$\Delta (\ln V_L) = a_0 L^{\omega(d=3)}+a_1$ that yields the value
\begin{eqnarray}
\omega(d=3) \simeq 0.24 
\label{omegadp3}
\end{eqnarray}
in agreement with the measures
of the droplet exponent $\omega_{DP}(1+2) \simeq 0.244$
obtained for the directed polymer in a 
random medium of dimension $1+2$ in various Refs
\cite{Tan_For_Wol,Ala_etal,KimetAla,Mar_etal,DP3d_us}.

\section{ Statistics of renormalized hoppings at criticality }

\label{secvcriti}

\subsection{ Expected multifractal statistics} 

At criticality, the statistics of the two-point transmission
is multifractal \cite{janssen99,us_twopoints,us_manywires} :
the critical probability distribution of the two-point transmission $T_L$ 
takes the form
\begin{equation}
{\rm Prob}\left( T_L \sim L^{-\kappa}  \right) dT
\oppropto_{L \to \infty} L^{\Phi(\kappa) } d\kappa
\label{phikappa}
\end{equation}
where the multifractal spectrum $\Phi(\kappa)$ exist only for $\kappa \geq 0$
(as a consequence of the physical bound $T_L \leq 1$) and is related
to the singularity spectrum $f(\alpha)$ of eigenfunctions via
\begin{eqnarray}
\Phi(\kappa \geq 0) = 2 \left[ f( \alpha= d+ \frac{\kappa}{2}   ) -d \right]
\label{resphikappa}
\end{eqnarray}

At criticality the decay of the two-point transmission 
is directly related to the decay of the renormalized hopping via
 Eq. \ref{trgdecay}. As a consequence, what is known about 
the statistics of the two-point transmission at criticality
can be translated for the renormalized hoppings.
The probability distribution of the renormalized hopping $V_L$ at scale $L$
takes the form
\begin{equation}
{\rm Prob}\left( \vert V_L \vert \sim L^{-\rho}  \right) dV
\oppropto_{L \to \infty} L^{H(\rho) } d\rho
\label{hrho}
\end{equation}
where
\begin{eqnarray}
H(\rho \geq 0) = \Phi( 2 \rho) = 2 \left[ f( \alpha= d+ \rho   ) -d \right]
\label{reshrho}
\end{eqnarray}

In particular, the typical exponent $\rho_{typ}$ characterizing the typical decay
\begin{eqnarray}
\overline{ \ln V_L } \opsimeq_{L \to +\infty} - \rho_{typ} \ln L
\label{vtypcriti}
\end{eqnarray}
is related to the typical exponent $\kappa_{typ}$ of the two-point transmission
and to the typical exponent $\alpha_{typ}$ of the singularity spectrum via
\begin{eqnarray}
 \rho_{typ} = \frac{\kappa_{typ}}{2} = \alpha_{typ}- d
\label{resrhotyp}
\end{eqnarray}

\subsection{ Results for the cubic lattice in dimension $d=3$ }

\label{resmultif3d}

\begin{figure}[htbp]
 \includegraphics[height=6cm]{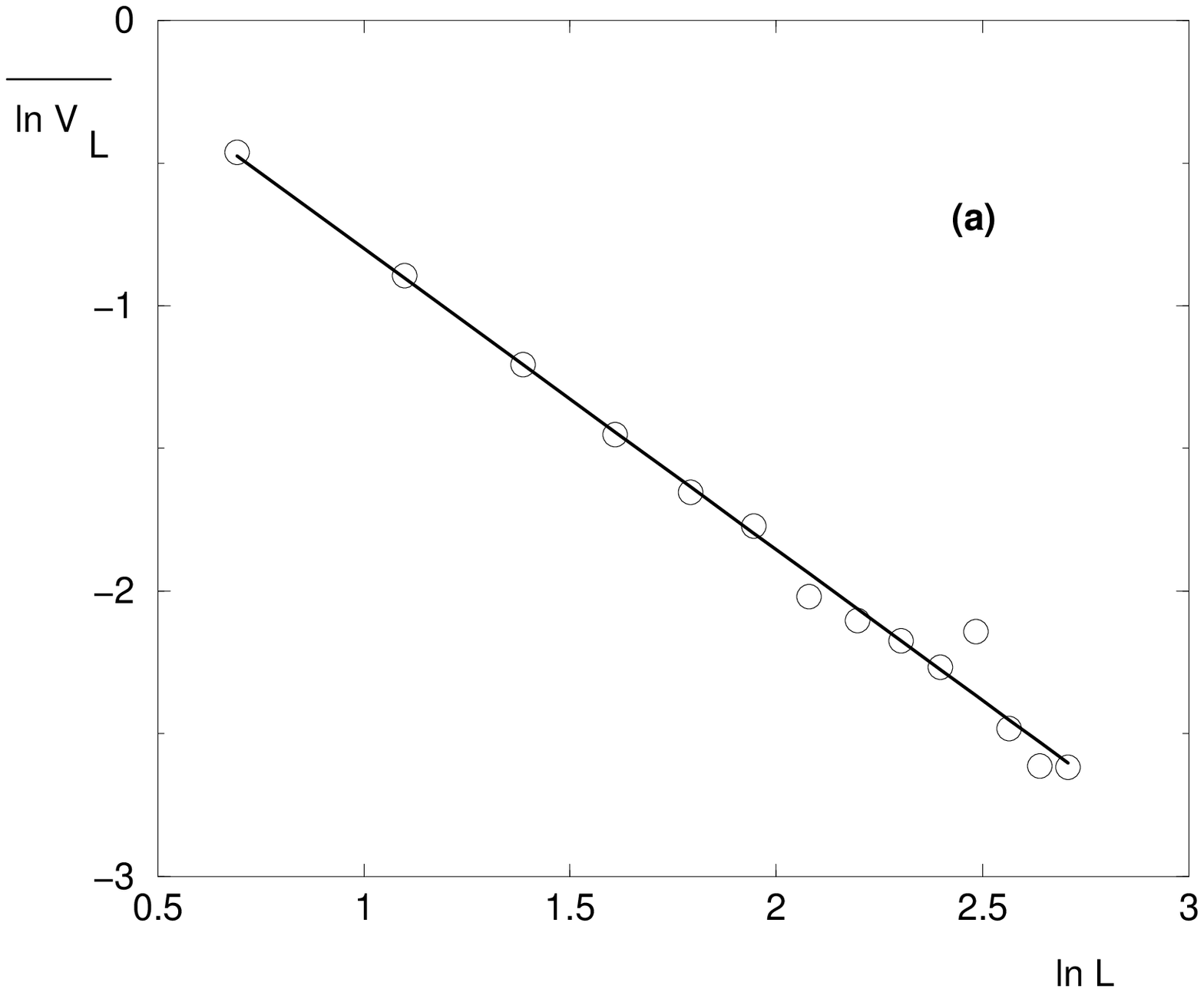}
\hspace{2cm}
 \includegraphics[height=6cm]{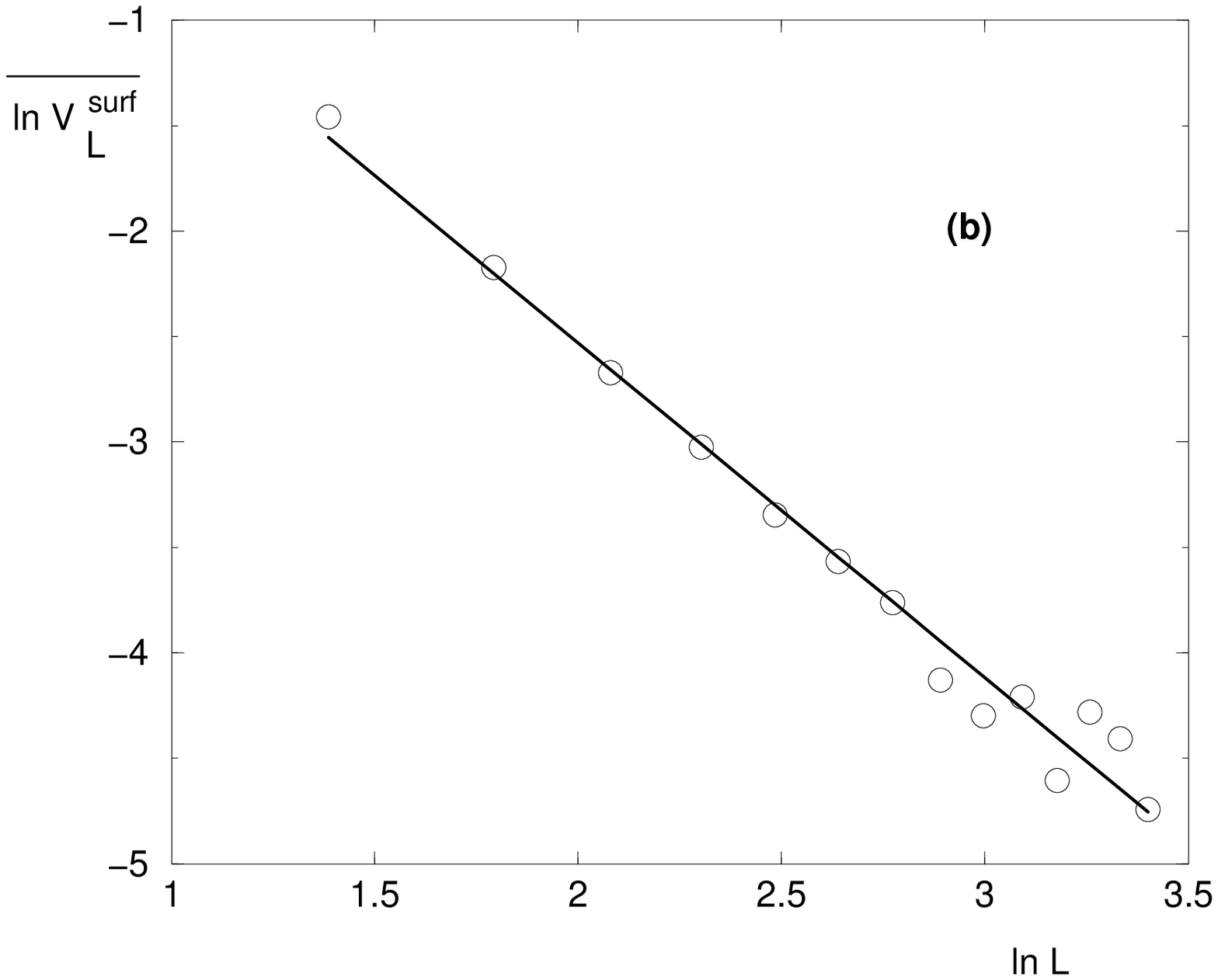}
\vspace{1cm}
\caption{
Statistics of the typical renormalized hopping
 in $d=3$ at criticality ($W=16.5$)
(a)  $\overline{ \ln V_L}$ as a function of $\ln L$
for bulk sites (see Fig. \ref{figrg3d}) :
the measured slope $\rho_{typ} \simeq 1.05$
represents the typical exponent of Eq. \ref{vtypcriti}.
(b) $\overline{ \ln V_L^{surf}}$ as a function of $\ln L$
for surface sites (see Fig. \ref{figrg3dsurf}) :
the measured slope $\rho_{typ}^{surf} \simeq 1.6 $ 
represents the typical exponent of Eq. \ref{vtypcritisurf}.  }
\label{fig3dvcriti}
\end{figure}

The typical exponent $\rho_{typ}$ of Eq. \ref{vtypcriti}
is measured from the data of Fig. \ref{fig3dvcriti} (a) 
\begin{eqnarray}
 \rho_{typ} \simeq 1.05
\label{rhotypnume}
\end{eqnarray}
Via Eq. \ref{resrhotyp}, this value is in agreement with the 
numerical measures of order \cite{mildenberger02,vasquez}
\begin{eqnarray}
 \alpha_{typ} \simeq 4
\label{alphatypnume}
\end{eqnarray}
for the exponent $\alpha_{typ}$
concerning the singularity spectrum of
eigenfunctions.  

Of course, beyond this typical exponent, one could in principle
extract from our numerical data, results on the whole multifractal spectrum.
However, our numerical means in $d=3$ are limited to rather
small sizes and small statistics (see the section \ref{numericaldetails})
in comparison with the exact diagonalization 
calculations of Refs \cite{vasquez}.
As a consequence, our numerical results seem sufficient to measure
the correct typical exponent, as shown above,
 but we believe that they are not sufficient
to measure correctly the rare events that are necessary to obtain 
a reliable multifractal spectrum. It may be that in the future,
more 'professional numericians' will be able to transform the
present renormalization approach into a competitive numerical method
to measure the multifractal spectrum, but this is clearly beyond
our numerical means.

\subsection{ Renormalized hopping between two surface points in dimension $d=3$ }

\begin{figure}[htbp]
 \includegraphics[height=6cm]{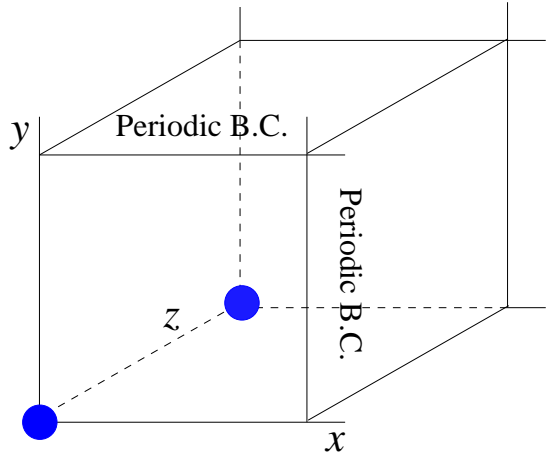}
\vspace{1cm}
\caption{ (Color on line) 
Renormalization procedure in dimension $d=3$ to measure 
the renormalized hopping $V_L^{surf}$ between two boundary sites.
The initial state is the tight-binding Anderson model
on a cubic lattice of size $L^3$,
 with periodic boundary conditions in two directions $x$ and $y$
and free boundary conditions in the third direction $z$.
Sites are then iteratively eliminated using the 
 the RG rules of Eqs. 
\ref{rulev} and \ref{rulee} until there remains only the two surface sites
corresponding to the large discs. }
\label{figrg3dsurf}
\end{figure}

At criticality, points lying on the boundaries are characterized
by a specific multifractal spectrum 
$f_{surf}(\alpha)$, different from
the bulk spectrum $f(\alpha)$ \cite{surf06,surfPRBM,surfhall}.
These surface critical properties are particularly interesting
in Anderson localization models where it is more natural to attach
leads to boundary sites rather than bulk sites.
We have thus considered the renormalization procedure depicted
on Fig. \ref{figrg3dsurf} to measure the statistical properties
of the renormalized hoppings between two surface points.
The typical behavior shown on Fig. \ref{fig3dvcriti} (b)
\begin{eqnarray}
\overline{ \ln V_L^{surf} } \opsimeq_{L \to +\infty} - \rho_{typ}^{surf} \ln L
\label{vtypcritisurf}
\end{eqnarray}
corresponds to an exponent of order 
\begin{eqnarray}
 \rho_{typ}^{surf}  \simeq 1.6
\label{rhotypsurfnume}
\end{eqnarray}
clearly distinct from its bulk analog of Eq. \ref{rhotypnume}.

Via Eq. \ref{resrhotyp}, we expect that this value corresponds to
\begin{eqnarray}
 \alpha_{typ}^{surf} = d+ \rho_{typ}^{surf} \simeq 4.6
\label{alphatypsurfnume}
\end{eqnarray}
for the typical exponent $\alpha_{typ}^{surf}$
of the surface singularity spectrum of eigenfunctions.
In contrast with the bulk case, we are not aware of any direct measure
of $\alpha_{typ}^{surf}$ 
in the literature to make some comparison.
As explained at the end of section \ref{resmultif3d},
we believe that our numerical means are not sufficient
to measure correctly the rare events to obtain the full 
multifractal spectrum around this typical value.
However, we expect that our result for the typical
exponent is reliable  (as shown above for the bulk case),
and will be confirmed in the future
 whenever the surface multifractal spectrum
will be measured via the powerful exact diagonalization techniques 
of \cite{vasquez}.

From Eq. \ref{resrhotyp}, we also expect that the two-point transmission
in $d=3$ between two surface points involves the typical exponent 
\begin{eqnarray}
 \kappa_{typ}^{surf} = 2 \rho_{typ}^{surf} \simeq 3.2
\label{kappatypnume}
\end{eqnarray}

\subsection{ Results for the PRBM model }

For the PRBM model, 
we have studied in detail the multifractal properties
of the two-point transmission  
in our previous works \cite{us_twopoints,us_manywires}.
Since the statistics of renormalized hoppings can be directly 
deduced from them via Eqs \ref{reshrho}, we refer the interested reader
to \cite{us_twopoints,us_manywires} where we have measured multifractal spectra
at criticality $a=1$ for various values of the parameter $b$.

\section{ Conclusion }

\label{secconclusion}

In this paper, we have revisited the exact real-space
renormalization procedure at fixed energy proposed by Aoki
 \cite{aoki80,aoki82,aokibook} for Anderson localization models.
We have presented detailed numerical results concerning
the statistical properties of the renormalized on-site energies $\epsilon$
 and of the renormalized hoppings $V$ as a function of the linear size $L$
for the Anderson tight-binding models in dimension $d=2$
where only the localized phase exists, and in 
dimension $d=3$ where there exists an Anderson localization transition.
Our main conclusions are the following :

(a)  the renormalized on-site energies
$\epsilon$ remain finite in the localized phase in $d=2,3$ and at
criticality ($d=3$), with a finite density at $\epsilon=0$ and a
power-law decay $1/\epsilon^2$ at large $\vert \epsilon \vert$.

(b) in the localized phase in dimension $d=2,3$,
the statistics of renormalized couplings belongs to the universality
class of the directed polymer in a
random medium of dimension $1+(d-1)$, in agreement with \cite{NSS,medina,prior}.
 
(c)  at criticality, the statistics of
renormalized hoppings $V$ is multifractal, in direct correspondence
with the multifractality of individual eigenstates and of two-point
transmissions. In particular, our measure $\rho_{typ}\simeq 1$
for the exponent governing the typical decay $\overline{{\rm ln} \ V_L} \simeq
-\rho_{typ} \ {\rm ln}L$, is in agreement with previous numerical measures
of $\alpha_{typ} =d+\rho_{typ} \simeq 4$ for the singularity spectrum $f(\alpha)$
of individual eigenfunctions. 
 We have also measured the corresponding critical surface properties.

In conclusion, we have shown that the large scale properties
of Anderson localization models actually emerge from the simple
real-space RG rules of Eqs \ref{rulev} and \ref{rulee} 
which preserve exactly the Green functions of the remaining sites.

\end{document}